%% file: Ignatev24_3.tex
\documentclass[fleqn,11pt]{article}
\usepackage{amsmath,amssymb}
\usepackage{xcolor}
\usepackage{verbatim}
\usepackage{dsfont}
\usepackage{graphicx}

\newtheorem{stat}{Statement}

\newtheorem{Mark}{Remark}
\newcommand{\Fig}[3]{%
\begin{center}
\parbox{8cm}{%
\refstepcounter{figure}\includegraphics[width=8cm,height=#2cm]{#1} \noindent Fig. \thefigure:\quad
#3}\end{center}}
\newcounter{strochka}

\newcounter{spisok}
\input preamble.tex

\begin{document}
\thispagestyle{empty}
\twocolumn[

\vspace{1cm}

\Title{Cosmological models based on an asymmetric scalar Higgs doublet with potential interactions between components}

\Author{Yu.G. Ignat'ev$^1$}
    {Institute of Physics, Kazan Federal University, Kremlyovskaya str., 16A, Kazan, 420008, Russia}

 \Author{A.R. Samigullina$^2$}
    {Institute of Physics, Kazan Federal University, Kremlyovskaya str., 16A, Kazan, 420008, Russia}

\Abstract
 {Cosmological models based on an asymmetric scalar Higgs doublet (canonical $\Phi$ and phantom $\phi$ fields) with potential interaction between the components are proposed. A qualitative analysis of the corresponding dynamic systems is performed and their transformation properties with respect to similarity transformations of fundamental constants are revealed. Asymptotic behavior of this class of cosmological models near cosmological singularities is investigated. Numerical simulations revealed a number of interesting features of these models, in particular, the possibility of a fairly long ``waiting phase'', during which the Universe is almost Euclidean, as well as the presence of bounce points, in the vicinity of which strong oscillations of scalar potentials occur.
\\[8pt]
}
\bigskip

] 
\section{Introduction}
Cosmological models based on the asymmetric scalar Higgs doublet (hereinafter, \emph{ASD}) were introduced in \cite{Yu_STFI_17}. These models were then partially studied in \cite{Yu_Kokh_Phys_1} -- \cite{Yu_Kokh_GC_2}. Finally, in \cite{Yu_Sam_Izv19} the behavior of the ASD model near the zero value of the Hubble parameter was investigated and a suggestion was made about the possibility of the existence of the so-called \emph{Euclidean cycles}, in which the total energy of the ASD tends to zero and the Universe is Euclidean. A significant limitation of the models studied in the above-mentioned works was the condition of non-negativity of the Hubble parameter $H(t)\geqslant0$, which did not allow a study of the full model. In \cite{Ignat_Dima_GC20}, this drawback was eliminated and a complete model for the case of the classical scalar Higgs singlet was presented, which allowed \cite{Yu_Kokh_TMF_21} to study in detail the cosmological models based on ASD, in which the classical $\Phi$ and phantom $\varphi$ components interact only via the gravitational field. Further, in \cite{Yu_Dima_TMF_21}, \cite{IgnatGC22}, based on the theoretical model of a statistical system of scalar charged fermions with the Higgs interaction potential, cosmological models were stu\-died in which the interaction between the compo\-nents of the \emph{ASD} was carried out via scalar charged fermions. In these papers, a collection of cosmolo\-gical models with scalar charged fermions, among which there are models with both a finite past (the initial singularity occurs at a finite time $t_0>-\infty$) and a finite future (BigRip, the final singularity occurs at a finite time $t_\infty<+\infty$). In \cite{IgnatGC22_2}, the behavior of cosmological models near these singu\-la\-ri\-ties $t\to t_s$ was investigated and the asymptotic value of the total ultrarelativistic equation of state of matter was found

\begin{equation}\label{kappa=1/3}
\kappa\equiv \frac{p}{\varepsilon}\to\frac{1}{3},\; (\mbox{at} \ t\to t_s).
\end{equation}
Further, in \cite{IgnatTMF24} the transformation properties of cosmological models with respect to similarity trans\-for\-ma\-tions of fundamental constants were studied, which allowed the results of numerical simulation to be extended to a wide class of models.

In connection with the problems of formation of supermassive black holes in the early Universe \cite{SMBH1e} -- \cite{Soliton} in the series of works \cite{Ignatev_SBH} -- \cite{TMF_23} a theory of scalar-gravitational instability of systems of scalar charged fermions in the ASD model was constructed, which satisfactorily explains the main characteristics of supermassive black holes. In this case, the interaction of scalar fields mediated through charged fermions gave a dynamic model of nonlinear interaction. Apparently, models of scalar fields with nonlinear interaction can play an important role in cosmology and astrophysics of the early Universe. The proposed model has two shortcomings: firstly, the extreme complexity of its mathematical structure, which complicates its analytical study, and, secondly, an insufficiently convincing physical argument for the appearance of scalar-charged fermions in the early stages of cosmological evolution.

In \cite{Yu_Kokh_24_1} the question was posed: is it possible to construct a cosmological model on a purely field basis, reflecting the main features of the model with scalar charge carriers? The answer to this question was devoted to the study of the ASD model with kinetic interaction of the components \cite{Yu_Kokh_24_1} -- \cite{Yu_Kokh_24_2}, mediated through the product of derivatives of scalar potentials. In this paper, we study the ADS model with potential interaction between the com\-po\-nents.

\section{Mathematical model\newline of cosmological evolution}
\subsection{Asymmetric scalar Higgs doublet with potential interaction between components}
Let us describe the basic relations of the mathematical model of cosmological evolution of the classical scalar Higgs field based on the asymmetric scalar doublet and its main properties. As a field model, we consider the self-consistent system of Einstein equations, classical $\Phi$ and phantom $\phi$ scalar fields, with the Higgs potential, which corresponds to the Lagrange function\footnote{Here and below, the Planck system of units $G=\hbar=c=1$ is used.}
\footnote{The work is performed according to the Russian Government Program of Competitive Growth of Kazan Federal University.}\\
\begin{eqnarray} \label{L}
L=\frac{1}{16\pi}(g^{lm} \Phi_{,l} \Phi_{,m}-2V(\Phi))\nonumber\\
+\frac{1}{16\pi}(-g^{lm}\varphi_{,l}\varphi_{,m}-2v(\varphi))
+\frac{1}{8\pi}\gamma\Phi^2\varphi^2,
\end{eqnarray}
where
\begin{equation} \label{V}
V(\Phi )=-\frac{\alpha }{4} \left(\Phi ^{2} -\frac{m^{2} }{\alpha } \right)^{2},
\end{equation}
\begin{equation} \label{v}
v(\varphi )=-\frac{\beta }{4} \left(\varphi ^{2} -\frac{\mu^{2} }{\beta } \right)^{2}
\end{equation}
–- the Higgs potential energy of the corresponding scalar fields, $\alpha$ and $\beta$ are their self-action constants, $\gamma$ is the interaction constant of the doublet components, $m$ and $\mu$ are their quantum masses.

The canonical energy-momentum tensor (see, for example, \cite{Land_Field}) of the scalar doublet with respect to the Lagrange function \eqref{L} has the form
\begin{eqnarray} \label{TEI}
8\pi T^k_i=\Phi_{,i}\Phi^{,k}-\frac{\delta^k_i}{2}\Phi_{,j}\Phi^{,j}+\delta^k_iV(\Phi)\nonumber\\
-\phi_{,i}\phi^{,k}+\frac{\delta^k_i}{2}\Phi_{,j}\Phi^{,j}+\delta^k_i v(\phi)
- \delta^k_i\gamma\Phi^2\phi^2.
\end{eqnarray}
The gravitational field of a scalar doublet is described by Einstein's equations with the energy-momentum tensor $T^k_i$ \eqref{TEI} and \emph{a bare cosmological constant} $\Lambda_0$
\begin{equation} \label{EqEn}
G^k_i=R^k_i-\frac{1}{2}R\delta^k_i=8\pi T^k_i+\Lambda_0\delta^k_i.
\end{equation}
The initial value of the cosmological constant $\Lambda_0$ is related to its observed value $\Lambda$, obtained by removing the constant term in the potential energy, by the relation
\begin{equation} \label{Lambda}
\Lambda=\Lambda_0-\frac{m^4}{4\alpha}-\frac{\mu^4}{4\beta}.
\end{equation}
Further, the Lagrange functions \eqref{L} correspond to the equations of scalar fields:
\begin{equation} \label{EqClassic}
\Delta\Phi+V'_\Phi -2\gamma\Phi\phi^2=0;
\end{equation}
\begin{equation} \label{EqFantom}
-\Delta\phi+v'_\phi -2\gamma\Phi^2\phi=0,
\end{equation}
where
\begin{equation} \label{Delta}
\Delta=\frac{1}{\sqrt{-g}}\frac{\partial}{\partial x^i}\sqrt{-g}g^{ik}\frac{\partial}{\partial x^k}.
\end{equation}

\subsection{Complete system of equations for the Friedmann metric}

In the case of a spatially flat Friedmann metric
\begin{equation} \label{ds}
ds^2=dt^2-a^2(t)(dx^2+dy^2+dz^2)
\end{equation}
where $a(t)$ is a scale factor, and the scalar fields $\Phi(t)$ and $\phi(t)$, which are independent of the three-dimensional coordinates, the energy-momentum tensor of the scalar field takes on an isotropic structure
\begin{equation} \label{T}
T^i_k=(\varepsilon+p)\delta^i_4\delta^4_k-p\delta^i_k,
\end{equation}
where
\begin{eqnarray} \label{T44}
\varepsilon\equiv T^4_4 & \displaystyle =\frac{1}{8\pi} \left(\frac{1}{2}\dot{\Phi}^2+V(\Phi)\right)+\nonumber\\
 &  \displaystyle \frac{1}{8\pi} \left(-\frac{1}{2}\dot{\phi}^2+v(\phi)\right)-\frac{1}{8\pi}\gamma\Phi^2\phi^2;\\
\label{Taa}
p\equiv-T^\alpha_\alpha & \displaystyle =\frac{1}{8\pi} \left(\frac{1}{2}\dot{\Phi}^2-V(\Phi)\right)\nonumber\\
 & \displaystyle -\frac{1}{8\pi}\left(\frac{1}{2}\dot{\phi}^2+v(\phi)\right)+\frac{1}{8\pi}\gamma\Phi^2\phi^2,
\end{eqnarray}
so
\begin{equation} \label{DEqEn} \varepsilon + p=\frac{1}{8\pi}(\dot{\Phi}^2-\dot{\phi}^2), \end {equation}
where $\varepsilon$ is the energy density and p is the pressure of the cosmological system.

The equations of the fields of the scalar doublet \eqref{EqClassic} -- \eqref{EqFantom} in the metric \eqref{ds} take the form
\begin{eqnarray} \label{DinamEq15}
\ddot{\Phi}+3\frac{\dot{a}}{a}\dot{\Phi}+V'_\Phi-2\gamma\Phi\phi^2=0;\\ \label{DinamEq2}
-\ddot{\phi}-3\frac{\dot{a}}{a}\dot{\phi}+v'_\phi-2\gamma\Phi^2\phi=0.
\end{eqnarray}

It is easy to see that of all Einstein's equations \eqref{EqEn} only two are non-trivial:
\begin{eqnarray} \label{EqEn44}
^4_4: 3\frac{\dot{a}^2}{a^2}-\frac{1}{2}\dot{\Phi}^2-V(\Phi)+\frac{1}{2}\dot{\phi}^2-v(\phi)\hspace{0.8cm}\nonumber\\
+\gamma\Phi^2\phi^2-\Lambda_0=0;\\
%
\label{EqEnalpha}
^\alpha_\alpha:2\frac{\ddot{a}}{a}+\frac{\dot{a}^2}{a^2}+\frac{1}{2}\dot{\Phi}^2-V(\Phi)-\frac{1}{2}\dot{\phi}^2-v(\phi)\nonumber\\
+\gamma\Phi^2\phi^2-\Lambda_0=0.
\end{eqnarray}
Let us give these equations a more compact form, moving from the independent variable $a(t)$ to the variable $H(t)$ (see \cite{Ignat_Dima_GC20} and the comments contained therein). To do this, we differentiate Einstein's equation with respect to time \eqref{EqEn44}
\begin{eqnarray} \label{DinEq1}
\frac{6\ddot{a}\dot{a}}{a^2}-\frac{6\dot{a}^3}{a^3}-\ddot{\Phi}\dot{\Phi}
-V'_\Phi\dot{\Phi}+\ddot{\phi}\dot{\phi}-v'_\phi \dot{\phi}\nonumber\\
-2\gamma\Phi\dot{\Phi}\phi^2-2\gamma\Phi^2\dot{\phi}\phi=0.
\end{eqnarray}
Multiplying both sides of the field equations \eqref{DinamEq15} and \eqref{DinamEq2} by $\dot{\Phi}$ and $\dot{\phi}$ respectively and substituting the result into \eqref{EqEn44}, we obtain
\begin{eqnarray} \label{DinamEq6}
3\frac{\dot{a}}{a}\left(\frac{2\ddot{a}}{a}-\frac{2\dot{a}^2}{a^2}+\dot{\Phi}^2-\dot{\phi}^2\right)=0.
\end{eqnarray}

Further introducing the Hubble parameter
\begin{equation} \label{H}
H=\frac{\dot{a}}{a},
\end{equation}
let's rewrite \eqref{DinamEq6} in the form
\begin{eqnarray}
6H\left(\dot{H}+\frac{\dot{\Phi}^2}{2}-\frac{\dot{\phi}^2}{2}\right)=0 \Rightarrow \nonumber\\
\label{dotH=}
\dot{H}=-\frac{1}{2}\dot{\Phi}^2+\frac{1}{2}\dot{\phi}^2.
\end{eqnarray}
Note that the equation \eqref{dotH=} can be written in an equivalent form by adding the equations \eqref{EqEn44} and \eqref{EqEnalpha} and applying the substitution \eqref{H}
\begin{eqnarray} \label{HdotPhiLambda0}
\dot{H}=& \displaystyle -3H^2+\frac{m^2\Phi^2}{2}-\frac{\alpha\Phi^4}{4}\hspace{1cm} \nonumber\\
 &  \displaystyle +\frac{\mu^2\phi^2}{2}-\frac{\beta\phi^4}{4}-\gamma\Phi^2\phi^2+\Lambda.
\end{eqnarray}
\subsection{Normal system of dynamic equations}
The resulting system of equations can be given a normal form, i.e., represented as a system of ordinary differential equations of the first order resolved with respect to derivatives:
\begin{eqnarray} \label{DinEq01}
\dot{\Phi}=Z,  \\ \label{DinEq02}
\!\!\!\!\!\!\dot{Z}=-3HZ-m^2\Phi+\alpha\Phi^3+2\gamma\Phi\phi^2,\\ \label{DinEq03}
\dot{\phi}=z, \\ \label{DinEq04}
\!\!\!\!\!\!\dot{z}=-3Hz+\mu^2\phi-\beta\phi^3-2\gamma\Phi^2\phi, \\ \label{DinEq05}
\dot{H}=-\frac{Z^2}{2}+\frac{z^2}{2}.
\end{eqnarray}
Taking into account \eqref{HdotPhiLambda0}, this system can be written in a form more convenient for qualitative analysis:
\begin{eqnarray} \label{DinEq1}
\dot{\Phi}=Z, &(\equiv F_1); \\ \label{DinEq2}
\!\!\!\!\!\!\dot{Z}=-3HZ-m^2\Phi+\alpha\Phi^3+2\gamma\Phi\phi^2, &(\equiv F_2);\\ \label{DinEq3}
\dot{\phi}=z, &(\equiv F_3); \\ \label{DinEq4}
\!\!\!\!\!\!\dot{z}=-3Hz+\mu^2\phi-\beta\phi^3-2\gamma\Phi^2\phi, &(\equiv F_4);\\ \label{DinEq5}
\dot{H}=-3H^2+\frac{m^2\Phi^2}{2}-\frac{\alpha\Phi^4}{4}\nonumber\\
+\frac{\mu^2\phi^2}{2}-\frac{\beta\phi^4}{4}-\gamma\Phi^2\phi^2+\Lambda, &(\equiv F_5).
\end{eqnarray}

Thus, the normal system of dynamic equations \eqref{DinEq1}--\eqref{DinEq5} describes phase trajectories in the five-dimensional phase arithmetic space $\mathbb{R}^5=\{\Phi$,$Z$,$\phi$, $z,H\}$, where $\Sigma_\Phi\cap \Sigma_\phi=\mathbb{R}\equiv OH$ and $\Sigma_\Phi\cup \Sigma_\phi=\mathbb{R}^5$ $(\mathbb{R}^3=\{\Phi,Z,H\} \-- \Sigma_\Phi, \mathbb{R}^3=\{\phi,z,H\} \-- \Sigma_\phi)$. Each specific phase trajectory determined by the initial conditions in this phase space corresponds to a specific cosmological model. Similarly to \cite{Ignat_Dima_GC20} it can be shown that the left-hand side of the equation \eqref{EqEn44} is the first integral of the dynamic system \eqref{DinEq1}--\eqref{DinEq5}:
\begin{eqnarray} \label{SurfEnst}
3H^2-\frac{Z^2}{2}+\frac{\alpha\Phi^4}{4}-\frac{m^2\Phi^2}{2} \nonumber\\
+\frac{z^2}{2}+\frac{\beta\phi^4}{4}-\frac{\mu^2\phi^2}{2}+\gamma\Phi^2\phi^2-\Lambda=0.
\end{eqnarray}
The equation \eqref{SurfEnst} defines the \emph{Einstein-Higgs hyper\-sur\-face}\footnote{see \cite{Ignat_Dima_GC20}.} in the phase space of the dynamical system \eqref{DinEq1}--\eqref{DinEq5}, on which all phase trajectories of this system lie. On the other hand, for given initial values $\Phi(0)\equiv \Phi_0$, $\phi(0)\equiv \phi_0$, $Z(0)\equiv Z_0$ and $z(0)\equiv z_0$, the equation \eqref{SurfEnst} defines the initial value of the Hubble parameter $H(0)\equiv H_0$. Two symmetric solutions of this equation $H{\pm}(0)\equiv \pm H_0$ correspond to starting from the expansion state $(+)$, or from the compression state $(-)$.

The equation \eqref{SurfEnst} can be written in the form
\begin{equation} \label{EqSurfEnst_E}
3H^2-\mathcal{E}=0,
\end{equation}
where the non-negative quantity $\mathcal{E}\geqslant0$ is the \emph{effective energy of the system}:
\begin{eqnarray} \label{EqE}
\mathcal{E}\equiv\frac{\dot{\Phi}^2}{2}-\frac{\alpha\Phi^4}{4}+\frac{m^2\Phi^2}{2} \nonumber\\
-\frac{\dot{\phi}^2}{2}-\frac{\beta\phi^4}{4}+\frac{\mu^2\phi^2}{2}-\gamma\Phi^2\phi^2+\Lambda.
\end{eqnarray}

Let us note useful relations for determining the \emph{invariant cosmological acceleration} $\Omega$:
\begin{eqnarray} \label{EqOmega}
\Omega=\frac{a\ddot{a}}{\dot{a}^2}\equiv 1+\frac{\dot{H}}{H^2}=-\frac{1}{2}(1+3\mathfrak{\kappa}),
\end{eqnarray}
where
\begin{eqnarray} \label{BarotropeCoeff}
\mathfrak{\kappa}=\frac{\varepsilon}{p}\Rightarrow \mathfrak{\kappa}=-\frac{1}{3}(1+2\Omega)
\end{eqnarray}
-- \emph{effective barotropic coefficient}.
\subsection{Dynamic system with scale factor}
$\mathbb{R}_6=\{\xi,\Phi,Z,\phi,z,H\}$
\begin{eqnarray} \label{EQScaleFactor}
	\dot{\xi}=H, &(\equiv F^*_1); \\ \label{EQSF1}
	\dot{\Phi}=Z, &(\equiv F^*_2); \\ \label{EQSF2}
	\!\!\!\!\!\!\dot{Z}=-3HZ-m^2\Phi+\alpha\Phi^3+2\gamma\Phi\phi^2, &(\equiv F^*_3);\\ \label{EQSF3}
	\dot{\phi}=z, &(\equiv F^*_4); \\ \label{EQSF4}
	\!\!\!\!\!\!\dot{z}=-3Hz+\mu^2\phi-\beta\phi^3-2\gamma\Phi^2\phi, &(\equiv F^*_5);\\ \label{EQSF5}
	\dot{H}=-3H^2+\frac{m^2\Phi^2}{2}-\frac{\alpha\Phi^4}{4}\nonumber\\
	+\frac{\mu^2\phi^2}{2}-\frac{\beta\phi^4}{4}-\gamma\Phi^2\phi^2+\Lambda, &(\equiv F^*_6) \label{EQSF6}.
\end{eqnarray}
\subsection{Symmetries}
Properties of symmetry of a dynamic system \eqref{DinEq1}--\eqref{DinEq5} -- its invariance under transformations
{\small
\begin{eqnarray} \label{Invariant}
\!\!\!\!\!\!\!\!\{\tau\!\rightarrow -\tau,\Phi\!\rightarrow\Phi,Z\!\rightarrow -Z,\phi\!\rightarrow\phi,z\!\rightarrow -z,H\!\rightarrow -H\};\\
\!\!\!\!\!\!\!\!\{\tau\!\rightarrow -\tau,\Phi\!\rightarrow -\Phi,Z\!\rightarrow Z,\phi\!\rightarrow -\phi,z\!\rightarrow z,H\!\rightarrow -H\}; \\
\!\!\!\!\!\!\!\!\{\tau\!\rightarrow \tau,\Phi\!\rightarrow -\Phi,Z\!\rightarrow -Z,\phi\!\rightarrow -\phi,z\!\rightarrow -z,H\!\rightarrow H\} .
\end{eqnarray}
}
\subsection{Scaling transformations\newline of the cosmological model}
It is easy to verify by analogy with \cite{IgnatTMF24} the invariance of the dynamic system \eqref{DinEq1}--\eqref{DinEq5}, as well as the equation \eqref{SurfEnst} with respect to the similarity trans\-for\-ma\-tion:
\begin{eqnarray} \label{Similar}
\alpha=k^2\tilde{\alpha},\beta=k^2\tilde{\beta},m=k\tilde{m},\mu=k\tilde{\mu}, \nonumber \\
t=k^{-1}\tilde{t}, \gamma=k^2\tilde{\gamma},\Lambda=k^2\tilde{\Lambda},\nonumber \\
\Phi=\tilde{\Phi},Z=k\tilde{Z}, \phi=\tilde{\phi},z=k\tilde{z},H=k\tilde{H}.
\end{eqnarray}
The following statement is true (see \cite{IgnatTMF24}):

\begin{stat} \label{stat_inv}
The scaling transformations \eqref{Similar} leave the system \eqref{DinEq1}--\eqref{DinEq5} invariant.
\end{stat}

Note that both the invariant acceleration \eqref{EqOmega} and the barotropic coefficient \eqref{BarotropeCoeff} are also invariant with respect to the transformation \eqref{Similar}
\begin{eqnarray} \label{SimilarOmegaSigma}
\Omega(t)=\tilde{\Omega}(kt), & \mathfrak{\kappa}(t)=\tilde{\mathfrak{\kappa}}(kt),
\end{eqnarray}
and the expression for the effective energy \eqref{EqE} is transformed according to the law
\begin{eqnarray} \label{SimilarOmegaSigma}
\mathcal{E}\rightarrow k^2\tilde{\mathcal{E}}.
\end{eqnarray}
Note (see \cite{IgnatTMF24}) that the equation \eqref{SurfEnst} of the Higgs hypersurface, which determines the topology of phase trajectories, is invariant with respect to the scale transformations \eqref{Similar}. Therefore, the topology of this surface does not change under the scale transformations \eqref{Similar}.

\begin{Mark} \label{Zam}In what follows, for brevity, we will denote the above-considered cosmological models based on an asymmetric scalar doublet with scalar interaction of components by the abbreviation ASD(P), and the models with kinetic interaction of components by ASD(K).
\end{Mark}

\section{Qualitative analysis\newline of a dynamic system}
\subsection{Singular points of a dynamic \newline system}
Singular points of a dynamic system \eqref{DinEq1}--\eqref{DinEq5} are determined by the equality to zero of the right-hand sides of these equations, i.e., by algebraic equations $F_i=0, (i=\overline{1..5})$:
\begin{eqnarray} \label{MS1}
Z=0,\ z=0;\\ \label{MS2}
-m^2\Phi+\alpha\Phi^3+2\gamma\Phi\phi^2=0;\\ \label{MS3}
\mu^2\phi-\beta\phi^3-2\gamma\Phi^2\phi=0;\\ \label{MS4}
-3H^2+\frac{m^2\Phi^2}{2}-\frac{\alpha\Phi^4}{4}\nonumber\\
+\frac{\mu^2\phi^2}{2}-\frac{\beta\phi^4}{4}-\gamma\Phi^2\phi^2+\Lambda=0.\label{MS5}
\end{eqnarray}
We multiply equation \eqref{MS2} by $\Phi$, and equation \eqref{MS3} by $\phi$ and make the substitution:
\begin{eqnarray} \label{EqChange}
\Phi^2\equiv x>0,&m^2x-\alpha x^2-2\gamma xy=0;\nonumber\\
\phi^2\equiv y>0,&\mu^2y-\beta y^2-2\gamma xy=0.
\end{eqnarray}
Then, in addition to the known solutions of the ASD model without interaction of components \cite{Yu_Kokh_TMF_21}--
\begin{eqnarray} \label{EqSolve}
 \displaystyle \{\Phi=0, \phi=0\}; & \nonumber\\
\left\{\Phi=0,\phi=\pm\frac{\mu}{\sqrt{\beta}}\right\}; &\!\!\! \displaystyle \left\{\Phi=\pm\frac{m}{\sqrt{\alpha}},\phi=0\right\}
\end{eqnarray}
if the condition is met
\begin{equation}\label{ab-4g2-not0}
\alpha\beta-4\gamma^2\neq 0
\end{equation}
we obtain, taking into account \eqref{EqChange}, four more solutions (taking into account the different values of the signs of the square roots)
\begin{eqnarray} \label{EqChange2}
\Phi^2=\displaystyle\frac{\beta m^2-2\gamma\mu^2}{\alpha\beta-4\gamma^2},& \phi^2=\displaystyle\frac{\alpha \mu^2-2\gamma m^2}{\alpha\beta-4\gamma^2}.
\end{eqnarray}
In this case, according to \eqref{SurfEnst}, the value of the square of the Hubble parameter at singular points is determined by the expression
\begin{eqnarray}
\label{Eq2}
H^2=\frac{1}{3}\Lambda+\frac{1}{12}\frac{\alpha\mu^4+\beta m^4-4\gamma m^2\mu^2}{\alpha\beta-4\gamma^2}.
\end{eqnarray}

Taking into account \eqref{EqChange2}, we can show the validity of the inequality
\begin{eqnarray}\label{H2=Lambda3}
H^2\geqslant\frac{\Lambda}{3}.
\end{eqnarray}

Thus, the coordinates of the singular points at $\gamma\neq 0$ and \emph{simultaneously} non-zero $\Phi$ and $\phi$ do not coincide with the coordinates of the corresponding points of the model without interaction of the components ($\gamma\equiv 0$) \cite{Yu_Kokh_TMF_21}.

As a result, 4 cases of solving the system of equations \eqref{MS1} -- \eqref{MS5} are possible:

1. Two symmetric points $M_{0,0}^\pm$ with zero potential and its derivative:
\begin{equation} \label{M00}
M_{0,0}^\pm: \left(0,0,0,0,\pm\displaystyle\frac{\sqrt{3\Lambda}}{3}\right)$, $\mathrm{if}\;  \Lambda\geqslant 0, \forall  \alpha, \beta.
\end{equation}
2. Four symmetric points $M_{\pm,0}^\pm$ located at the vertices of a rectangle in the plane $\Phi$ and $H$:
\begin{eqnarray} \label{Mpm00}
M_{\pm,0}^\pm: \left(\displaystyle\pm\frac{m}{\sqrt{\alpha}},0,0,0,\pm\displaystyle\sqrt{\frac{m^4+4\alpha\Lambda}{12\alpha}}\right), \nonumber \\
\mathrm{if} \; \displaystyle\frac{m^4}{4\alpha}+\Lambda\geqslant 0, \forall\; \alpha>0, \; \beta.
\end{eqnarray}
3. Four symmetric points $M_{0,\pm}^\pm$ located at the vertices of a rectangle in the plane $\phi$ and $H$:
\begin{eqnarray} \label{M00pm}
M_{0,\pm}^\pm: \left(0,0,\displaystyle\pm\frac{\mu}{\sqrt{\beta}},0,\pm\displaystyle\sqrt{\frac{\mu^4+4\beta\Lambda}{12\beta}}\right), \nonumber \\
\mathrm{if} \; \displaystyle\frac{\mu^4}{4\beta}+\Lambda\geqslant 0, \forall\;\alpha, \beta>0.
\end{eqnarray}
4. Four symmetric points $M_{\pm,\pm}^\pm$ located at the vertices of the rectangle in the plane $\Phi$ and $H$ and 4 symmetric points $M_{\pm,\pm}^\pm$ located at the vertices of the rectangle in the plane $\phi$ and $H$:
{\small
\begin{eqnarray} \label{Mpmpm}
\!\!\!\!\!\!M_{\pm,\pm}^\pm:\!\!\Biggl(\!\!\displaystyle\pm\sqrt{\frac{\beta m^2-2\gamma\mu^2}{\alpha\beta-4\gamma^2}},0,
\displaystyle\pm \sqrt{\frac{\alpha\mu^2-2\gamma m^2}{\alpha\beta-4\gamma^2}},\nonumber\\
0,\pm\displaystyle\frac{1}{\sqrt{3}}\sqrt{\Lambda+\frac{\alpha\mu^4+\beta m^4-4\gamma m^2\mu^2}{4(\alpha\beta-4\gamma^2)}}\Biggr),
\nonumber \\
\mathrm{if} \; H^2\geqslant\frac{\Lambda}{3} \; \mathrm{then} \; \forall \; \alpha>2\gamma \frac{m^2}{\mu^2},\;\beta>2\gamma \frac{\mu^2}{m^2}.
\end{eqnarray}
}

As can be seen from the previous formulas, the interaction constant $\gamma$ affects only the coordinates of the singular points of $M_{\pm,\pm}^\pm$. In the absence of interaction ($\gamma=0$), the condition for the existence of these points is
\begin{eqnarray}\label{noeq_gamma}
4\Lambda\alpha\beta+\alpha\mu^4+\beta m^4\geqslant 0,
\end{eqnarray}
which is identically satisfied for $\Lambda\geqslant 0$. Let us find out when such points can exist in our model for $\gamma\not\equiv 0$.
For the existence of singular points of $M_{\pm,\pm}^\pm$, the condition must be satisfied
\begin{eqnarray}\label{noeq_gamma}
\Lambda+\frac{\alpha\mu^4+\beta m^4-4\gamma m^2\mu^2}{4(\alpha\beta-4\gamma^2)}\geqslant0.
\end{eqnarray}
Thus, we get
\begin{eqnarray}
\gamma^2<\frac{1}{4}\alpha\beta&\Leftrightarrow
\Lambda\gamma^2+\frac{1}{4}\gamma m^2\mu^2\nonumber \\
&-\frac{1}{4}\Lambda\alpha\beta-\frac{1}{16}(\alpha\mu^4+\beta m^4)\leqslant 0;\nonumber \\
\gamma^2>\frac{1}{4}\alpha\beta&\Leftrightarrow
\Lambda\gamma^2+\frac{1}{4}\gamma m^2\mu^2\nonumber \\
&-\frac{1}{4}\Lambda\alpha\beta-\frac{1}{16}(\alpha\mu^4+\beta m^4)\geqslant 0.\nonumber
\end{eqnarray}
Let us further introduce the roots of the equation
\begin{eqnarray}\label{eq_gamma}
\!\!\!\Lambda\gamma^2+\frac{1}{4}\gamma m^2\mu^2-\frac{1}{4}\Lambda\alpha\beta-\frac{1}{16}(\alpha\mu^4+\beta m^4)=0 \rightarrow\nonumber
\end{eqnarray}
\begin{eqnarray}\label{gamma=}
\gamma_\pm=& \displaystyle \frac{1}{8\Lambda}\biggl[-m^2\mu^2\pm\bigl(16\Lambda^2\alpha\beta \nonumber\\
 & \displaystyle +m^4\mu^4+4\Lambda(\alpha\mu^4+\beta m^4)\bigr)^{\frac{1}{2}}\biggr].
\end{eqnarray}
Thus, the following statement is true.
\begin{stat}\label{stat1}
The condition for the existence of singular points $M_{\pm,\pm}^\pm,$ i.e., the solution of the ine\-qua\-lity \eqref{noeq_gamma} is:
\begin{eqnarray}\label{Lambda>0}
\Lambda>0&\Rightarrow&\gamma\in \biggl(-\frac{1}{2}\sqrt{\alpha\beta},\frac{1}{2}\sqrt{\alpha\beta}\biggr);\\ \label{Lambda<0}
\Lambda<0&\Rightarrow&\gamma\in (-\infty,\gamma_-]\cup[\gamma_+,+\infty).
\end{eqnarray}
If the conditions \eqref{Lambda>0} or \eqref{Lambda<0} are not met, there are no singular points $M_{\pm,\pm}^\pm$.
\end{stat}

\subsection{Eigenvalues of the matrix of a dynamic system}
The characteristic matrix $\mathbf{A(M)}=||A^k_i|| \equiv\left\|\partial F_{i}/\partial x_{k} \right\|$\footnote{see, for example, \cite{Bogoyav}.} of a dynamic system \eqref{DinEq1}--\eqref{DinEq5} at an arbitrary point $\mathbf{M}$ of the phase space $\mathbb{R}^5$ has the following form:
{\small
\begin{equation}\label{AM}
\!\!\!\!\!\!\!\!\! \mathbf{A(M)}=\left(\begin{array}{ccccc}
\vspace{2mm}
\!\!\! 0&1&0&0&0\\
\vspace{2mm}
\!\!\! \displaystyle\frac{\partial F_2}{\partial \Phi}&-3H&\displaystyle\gamma \frac{\partial F_2}{\partial \phi}&0&-3Z\\
\vspace{2mm}
\!\!\! 0&0&0&1&0\\
\vspace{2mm}
\!\!\! \displaystyle-\gamma \frac{\partial F_4}{\partial \Phi}&0&\displaystyle\frac{\partial F_4}{\partial \phi}&-3H&-3z\\
\vspace{2mm}
\!\!\!\displaystyle\frac{\partial F_5}{\partial \Phi}&0&\displaystyle\frac{\partial F_5}{\partial \phi}&0&-6H
	\end{array}
	\right)\!\!.\!\!\!\!
\end{equation}}
Calculating the matrix $\mathbf{A(M)}$ of the dynamic system \eqref{DinEq1}--\eqref{DinEq5} at its singular points $\mathbf{M_0}$, we find, taking into account \eqref{MS1}--\eqref{MS4}
{\small
\begin{equation}\label{AM0}
\!\!\!\!\!\!\! \mathbf{A(M_0)}=\!\!\left(\begin{array}{ccccc}
\!\!\! 0&1&0&0&0\\
\!\!\! 2\alpha\Phi^2 &-3H &4\gamma\Phi\phi &0&0\\
\!\!\! 0&0&0&1&0\\
\!\!\! -4\gamma\Phi\phi&0&-2\beta\phi^2&-3H&0\\
\!\!\! 0&0&0&0&-6H
\end{array}\!\!\!\!
\right)\!\!.\!\!
\end{equation}}
The determinant of the matrix \eqref{AM0},
\begin{equation}\label{DetAM0}
\mathrm{det}(\mathbf{A(M_0)})=24\Phi^2\phi^2H(\alpha\beta-4\gamma^2),
\end{equation}
when the condition \eqref{ab-4g2-not0} is satisfied, vanishes in the cases $\Phi=0$ or $\phi=0$, or $H=0$.
The characteristic equation  fot the eigenvalues $\lambda$ with respect to the matrix \eqref{AM0} takes the form:
{\small
\begin{equation}
	\begin{array}{l}\label{CharAM0}
\!\!\!\!\!\!\!(\lambda + 6H)\times[-4\alpha\Phi^2\beta\phi^2 +16\gamma^2\Phi^2\phi^2+ 9H^2\lambda^2+ 6H\lambda^3 \nonumber \\[12pt]
-6\alpha H\Phi^2\lambda+6H\beta\lambda\phi^2 -2\alpha\Phi^2\lambda^2 + 2\beta\lambda^2\phi^2 +\lambda^4]=0.
\end{array}
\end{equation}}

Thus, the eigenvalues of the matrix \eqref{AM0} of the dynamic system \eqref{DinEq1}--\eqref{DinEq5} are equal to:
\begin{eqnarray}
	\label{lambdas}
		\lambda_i=\displaystyle-\frac{3}{2}H\pm \displaystyle\frac{1}{2}\sqrt{a\pm 4\sqrt{b}},& (i=\overline{1,4}) ,\\ & \lambda_5=-6H,
	\end{eqnarray}
where
\begin{eqnarray}
a\equiv 9H^2 + 4\alpha\Phi^2-4\beta\phi^2,\nonumber\\
b\equiv\alpha^2\Phi^4+2\alpha\Phi^2\beta\phi^2-16\gamma^2\Phi^2\phi^2+\beta^2\phi^4,\nonumber
\end{eqnarray}
where it is necessary to substitute the values of the functions $\Phi$ and $\phi$ at the singular points \eqref{M00} -- \eqref{Mpmpm}.
\subsection{Classification of non-degenerate\newline singular points}

As a result, we obtain the following classification of singular points (see Table \ref{TabPoint1}), where the character of the points is indicated by an ordered list of the type $[\mathbf{A},\mathbf{S}]$, respectively, to the phase subspaces $\Sigma_\Phi$ and $\Sigma_\phi$, where \textbf{A} denotes the attractive point, \textbf{R} the repulsive point, and \textbf{S} the saddle point.

\begin{center}
	\refstepcounter{table}
	{\bf Tab \thetable.} \label{TabPoint1}\hskip 12ptClassification of non-degenerate singular points $\Sigma_\Phi$, $\Sigma_\phi$ \\[12pt]
	\begin{tabular}[c]{|c|c|}
\hline
Point &  Type\\ \hline
$M_{0,0}^{+}$ &  [\textbf{A}, \textbf{S}] \\ \hline
$M_{0,0}^{-}$ &  [\textbf{R}, \textbf{S}] \\ \hline
$M_{\pm,0}^{+}$ & [\textbf{S}, \textbf{S}] \\ \hline
$M_{\pm,0}^{-}$ &  [\textbf{S}, \textbf{S}] \\ \hline
$M_{0,\pm}^{+}$ & [\textbf{A}, \textbf{A}] \\ \hline
$M_{0,\pm}^{-}$ &  [\textbf{R}, \textbf{R}] \\ \hline
$M_{\pm,\pm}^{+}$ &  [\textbf{S}, \textbf{A}] \\ \hline
$M_{\pm,\pm}^{-}$ &  [\textbf{S}, \textbf{R}] \\ \hline
\end{tabular}
\label{tab1}
\end{center}
Thus, the following statement is true.

\begin{stat}\label{stat_stable}
\hspace{-6pt}. For
\[\mu^4+4\beta\Lambda>0\]
the dynamical system \eqref{DinEq1}--\eqref{DinEq5} has two and only two stable (attractive) singular points $M_{0,\pm}^{+}$ \eqref{M00pm}.
\end{stat}

\section{Asymptotic properties of the model near the singularity}

By analogy with \cite{IgnatGC22_2} and \cite{Yu_Kokh_24_1}, we consider the behavior of the cosmological model near the cosmological singularity $a=0$, $\xi \rightarrow - \infty$.
The beginning of the Universe (the cosmological singularity $a=0$) corresponds to $\xi \rightarrow - \infty$, and the infinite future $a \rightarrow \infty$ -- $\xi \rightarrow + \infty$.

In order to proceed to the consideration of the model near the cosmological singularity, we define the relationships between the field quantities from the equation \eqref{SurfEnst} near $t\rightarrow t_s$:
{\small
\begin{eqnarray}	\label{PhiZphizH}
\!\!\!\!\!\!\mid \Phi \mid\rightarrow\infty, \mid Z \mid\rightarrow\infty, \mid \phi \mid\rightarrow\infty, \mid z \mid\rightarrow\infty, \mid H \mid\rightarrow\infty; \nonumber\\
\!\!\!\!\! \Phi^4\sim \phi^4 \lesssim Z^2\sim z^2 \sim H^2,\; \Phi^2\sim\phi^2\lesssim \mid Z\mid \sim \mid H\mid.
\end{eqnarray}
}
According to the asymptotic formulas \eqref{PhiZphizH} near the cosmological singularity, the system of equations \eqref{DinEq01} -- \eqref{DinEq05} will take the following form:\\
\begin{eqnarray}
\label{EQSF60}
\dot{\xi}=H;\\
\label{EQSF61}
\dot{\Phi}=Z; \; \dot{\phi}=z;\\
\label{EQSF62}
\dot{Z}=-3HZ; \; \dot{z}=-3Hz;\\
\label{dotH=Z}
\dot{H}=-\frac{1}{2}Z^2+\frac{1}{2}z^2.
\end{eqnarray}
Using \eqref{EQSF60} we will replace variables $t\to\xi$
\begin{equation}\label{t->xi}
\frac{df}{dt}=\frac{df}{d\xi}H\Rightarrow \dot{f}=f'H
\end{equation}
in the system of equations \eqref{EQSF61} -- \eqref{dotH=Z}:
\begin{eqnarray}
\label{EQSF61_xi}
\Phi'=\frac{Z}{H};\quad \phi'=\frac{z}{H};\\
\label{EQSF62_xi}
Z'=-3Z;\quad z'=-3z;\\
\label{EQSF63_xi}
\label{H'=Z}
(H^2)'=-Z^2+z^2.
\end{eqnarray}
Integrating \eqref{EQSF62_xi}, we find:
\begin{equation}\label{Z,z=}
Z=C_1 e^{-3\xi};\quad z=C_2 e^{-3\xi},
\end{equation}
where $C_1,C_2$ are arbitrary constants. Substituting \eqref{Z,z=} into the equation \eqref{dotH=Z} and performing elementary integration, we find its solution
\begin{eqnarray}\label{h^2=}
H^2= \frac{1}{6}(C_1^2-C_2^2)e^{-6\xi}\Rightarrow\nonumber\\
H=\pm \frac{e^{-3\xi}}{\sqrt{6}}\sqrt{C_1^2-C_2^2},\quad (C_1^2-C_2^2\geqslant0) .
\end{eqnarray}
We have discarded the additive integration constant $C_3$ due to its smallness in comparison with the singular terms. Thus, taking into account \eqref{dotH=Z} and \eqref{h^2=} we find
\begin{eqnarray}\label{dotH=xi}
\dot{H}=-\frac{1}{2}(C_1^2-C_2^2)e^{-6\xi}\Rightarrow\\
\Omega=1+\frac{\dot{H}}{H^2}\Rightarrow \Omega=-2\Rightarrow \kappa=1.
\end{eqnarray}

Substituting the found solutions \eqref{Z,z=} and \eqref{dotH=xi} into the equations \eqref{EQSF61_xi} and performing integration, we obtain asymptotic solutions for the potentials $\Phi$ and $\phi$:
\begin{eqnarray}	\label{AsimpPhi}
\Phi=\pm\frac{\sqrt{6}C_1}{\sqrt{C_1^2-C_2^2}}\xi ;
\end{eqnarray}
\begin{eqnarray}	\label{Asimpphi}
\phi=\pm\frac{\sqrt{6}C_2}{\sqrt{C_1^2-C_2^2}}\xi.
\end{eqnarray}
Note that the obtained asymptotic solutions \eqref{Z,z=}, \eqref{dotH=xi}, \eqref{AsimpPhi} and \eqref{Asimpphi} confirm the asymptotic estimates \eqref{PhiZphizH}.

\section{Numerical Simulation}
\subsection{Preliminary Remarks}
Proceeding to numerical simulation, we first define a set of fundamental parameters of the model using an ordered list
\begin{eqnarray} \label{Param}
\mathbf{P}=[[\alpha,\beta,m,\mu],\Lambda,\gamma].
\end{eqnarray}
Secondly, we note that the dynamic system \eqref{DinEq01} -- \eqref{DinEq05} is autonomous, and therefore invariant with respect to the transformation $t\to t_0+t$. As a result, we can always choose $t_0$ so that
\begin{eqnarray}	\label{a=1}
a(0)=1 \Rightarrow \xi(0)=0.
\end{eqnarray}
Thirdly, we note that the Hubble parameter $H(t)$ for given $\Phi_0$,$Z_0$,$\phi_0$, $z_0$ is determined up to the sign by the Einstein equation \eqref{SurfEnst}, from which we find
\begin{eqnarray} \label{H0}
H_0=\pm\frac{1}{\sqrt{3}}
\biggl(\frac{Z_0^2}{2}-\frac{\alpha\Phi_0^4}{4}+\frac{m^2\Phi_0^2}{2}\nonumber \\
-\frac{z_0^2}{2}-\frac{\beta\phi_0^4}{4}+\frac{\mu^2\phi_0^2}{2}-\gamma\Phi_0^2\phi_0^2+\Lambda\displaystyle\biggr)^{1/2}.
\end{eqnarray}
Given \eqref{H0}, we will specify the initial conditions as an ordered list
\begin{eqnarray} \label{IC}
\mathbf{I}=[\Phi_0,Z_0,\phi_0,z_0,e],
\end{eqnarray}
where $e=\pm1$ is the indicator of the sign of the solution \eqref{H0}; in this case, the initial condition \eqref{a=1} is assumed to be satisfied.

Fourthly, we note that below we consider para\-meters with values close to 1 as the basic para\-meters of the model. To move to real physical para\-meters, it is necessary at the final stage of numerical modeling to perform a large-scale transformation of the para\-meters and functions using the model invariants \eqref{Similar}.

\subsection{Models with initial conditions at critical points}
Let's consider a group of models with parameters \eqref{par1.1}--\eqref{par1.3}
\begin{eqnarray}	\label{par1.1}
\mathbf{P^{(1)}_1}&=&[[1,1,1,1],1,1];\\ \label{par1.2}
\mathbf{P^{(1)}_2}&=&[[1,1,1,1],10^{-1},1];\\ \label{par1.3}
\mathbf{P^{(1)}_3}&=&[[1,1,1,1],0,1]
\end{eqnarray}
and initial conditions \eqref{ic11}, \eqref{ic12}
\begin{eqnarray}	\label{ic11}
\mathbf{I^{(1)}_{1\pm}}=[1,0,0,0,\pm 1],\\ \label{ic12}
\mathbf{I^{(1)}_{2\pm}}=[0,0,1,0,\pm 1].
\end{eqnarray}
Note that the initial conditions \eqref{ic11}, \eqref{ic12} coincide with the coordinates of the critical points
\begin{eqnarray}
\!\!\!\!\!M^\pm_{+,0}\!=[1,0,0,0,\pm H_0], M^\pm_{0,+}\!=[0,0,1,0,\pm H_0].\nonumber
\end{eqnarray}

Fig. \ref{fig1} shows the evolution of the geometric factors $\xi(t)$ and $H(t)$ in the models $\mathbf{P^{(1)}_1}$ \eqref{par1.1}, $\mathbf{P^{(1)}_2}$ \eqref{par1.2}, and $\mathbf{P^{(1)}_3}$ \eqref{par1.3} with initial conditions $\mathbf{I^{(1)}_{1\pm}}$ \eqref{ic11} and $\mathbf{I^{(1)}_{2\pm}}$ \eqref{ic12}.
\Fig{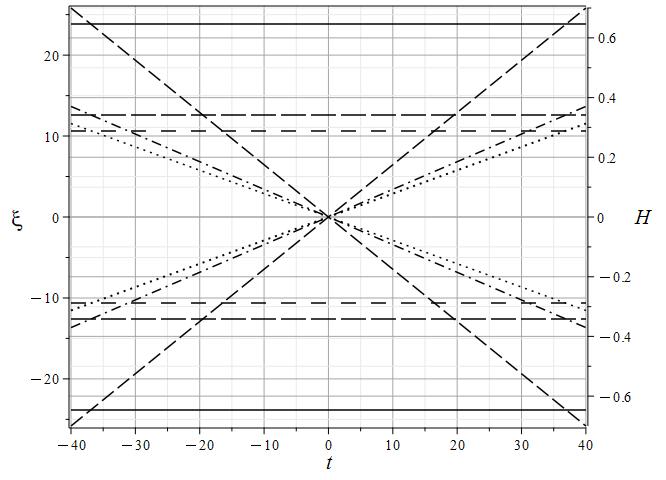}{6}{\label{fig1} $\xi(t)$: dashed line -- $\Lambda=1$; dash-dotted -- $\Lambda=0.1$; dotted line -- $\Lambda=0$ and $H(t)$: solid line -- $\Lambda=1$; long-dashed -- $\Lambda=0.1$; long-dotted -- $\Lambda=0$; $\mathbf{P^{(1)}_1}$ -- $\mathbf{P^{(1)}_3}$ and initial conditions $\mathbf{I^{(1)}_{1\pm}}$.}

Note that, as expected, the coordinates of the singular points of the model are exact solutions of the dynamic system of equations \eqref{DinEq1}--\eqref{DinEq5}. There\-fore, with equal parameters, these models are indis\-ti\-n\-gui\-shable from each other at the critical points $M^\pm_{\pm,0}$ and $M^\pm_{0,\pm}$. It is for this reason that Fig. \ref{fig1} shows the results of numerical integration for only one of these models.
\subsection{Models with initial conditions outside critical points}
\subsubsection{Model with infinite past and future with initial conditions near the singular point $M^+_{0,0}$; $\Lambda=0$ }
Consider the model with fundamental parameters
\begin{eqnarray} \label{par2}
\mathbf{P^{(2)}}&=&[[1,1,1,1],0,5\cdot 10^{-5}].
\end{eqnarray}
and initial conditions \eqref{ic2}
\begin{eqnarray}	\label{ic2}
\mathbf{I^{(2)}}=[10^{-3},0,10^{-5},0,1].
\end{eqnarray}
The characters of the special points for the \eqref{par2} model are presented in Table \ref{tab2}\footnote{Table \ref{tab2} -- \ref{TabPoint} presents the rounded values of the coordinates of the special points.}.
\begin{center}
	\refstepcounter{table}
	{\bf Tab \thetable.} \label{TabPoint}\hskip 12pt Special points of the model \eqref{par2} \\[12pt]
	\begin{tabular}[c]{|c|c|c|}
\hline
Point & Coordinates & Type\\ \hline
$M_{0,0}^{\pm}$ &[0, 0, 0, 0, 0]& [\textbf{S}, \textbf{S}] \\ \hline
$M_{\pm,0}^{+}$ &[$\pm$ 1, 0, 0, 0, 0.289]& [\textbf{S}, \textbf{S}] \\ \hline
$M_{\pm,0}^{-}$ &[$\pm$ 1, 0, 0, 0, -0.289] & [\textbf{S}, \textbf{S}] \\ \hline
$M_{0,\pm}^{+}$ &[ 0, 0,$\pm$ 1, 0, 0.289]& [\textbf{A}, \textbf{A}] \\ \hline
$M_{0,\pm}^{-}$ & [ 0, 0,$\pm$ 1, 0, -0.289] & [\textbf{R}, \textbf{R}] \\ \hline
$M_{\pm,\pm}^{+}$ &[$\pm$ 1, 0, $\pm$ 1, 0, 0.408]&  [\textbf{S}, \textbf{A}] \\ \hline
$M_{\pm,\pm}^{-}$ &[$\pm$ 1, 0, $\pm$ 1, 0,- 0.408]&  [\textbf{S}, \textbf{R}] \\ \hline
\end{tabular}
\label{tab2}
\end{center}

This model belongs to the class of cosmological models with infinite past and future.
Fig. \ref{fig2} shows the evolution of geometric factors $\xi(t)$ and $H(t)$, and Fig. \ref{fig3} shows the evolution of scalar potentials for the model with parameters $\mathbf{P^{(2)}}$ and initial conditions $\mathbf{I^{(2)}}$.
As can be seen from the graphs in Fig. \ref{fig2}, this model has an inflationary beginning and end, with an inflationary contraction at the initial stage with $H\approx-0.289$, which is replaced by an inflationary expansion at the final stage with $H\approx +0.289$. The above inflationary stages are connected by a ``bridge'' with a very small constant value of the Hubble parameter $H\approx 4\cdot10^{-4}$, $\xi\to 0\Rightarrow a(t)\approx 1$. We can say that on the interval $t\in [-15,15]$ the Universe is almost Euclidean. Since the total energy of the cosmological system $\mathcal{E}$ according to \eqref{EqSurfEnst_E} is equal to $3H^2$, then at this intermediate stage the total energy tends to zero $\mathcal{E}\approx 10^{-7}$. Note that the assumption about the existence of Euclidean cycles was formulated in the work of the authors \cite{Yu_Sam_Izv19}.

At the same time, strong oscillations of the scalar field potentials occur precisely at the intermediate stage: the potential of the classical field $\Phi$ leaves the stable state $\Phi=0$ and after several oscillations returns to its previous state. The potential of the phantom field $\phi$ passes from the stable state $\phi=1$ to the unstable state $\phi=0$ and then returns to its previous state.
\Fig{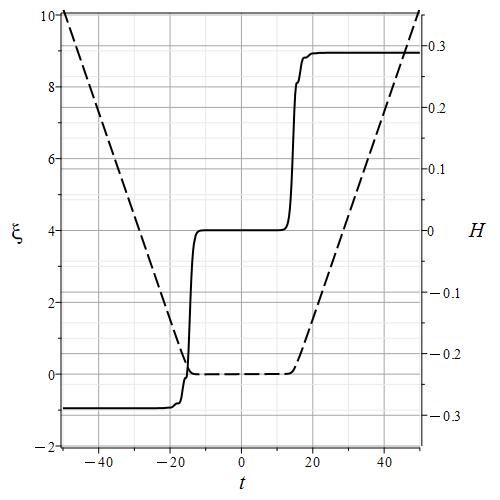}{6}{\label{fig2} $\xi(t)$ -- dashed line; $H(t)$ -- solid line.}
\Fig{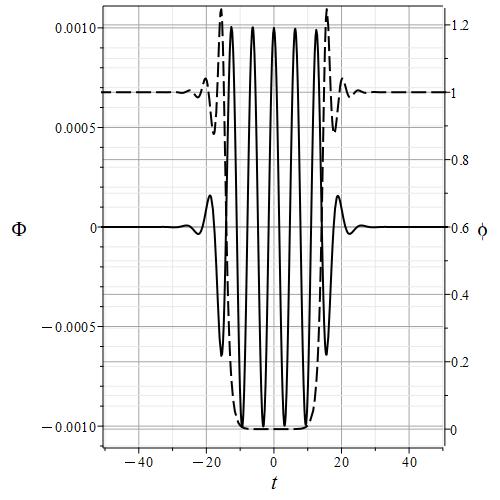}{6}{\label{fig3} $\Phi(t)$ -- solid line and phantom scalar field $\phi(t)$ -- dashed line.}
\Fig{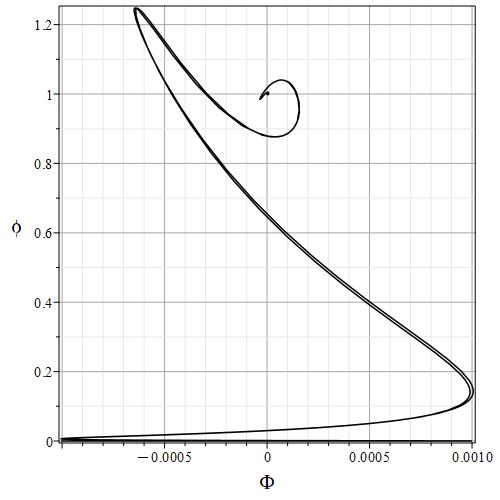}{7}{\label{fig4} Phase trajectory of the model $\mathbf{P^{(2)}}$ in the plane $\{\Phi,\phi\}$.}
Fig. \ref{fig4} -- \ref{fig5} show phase diagrams of this process -- two-dimensional in Fig. \ref{fig4} and three-dimensional in Fig. \ref{fig5}. These diagrams show the details of the transition of the dynamic system to the stable singular point $\Phi=0,\ \phi=1 $. The double line in Fig. \ref{fig4} is obtained by superimposing projections of different sections of the three-dimensional line.

\Fig{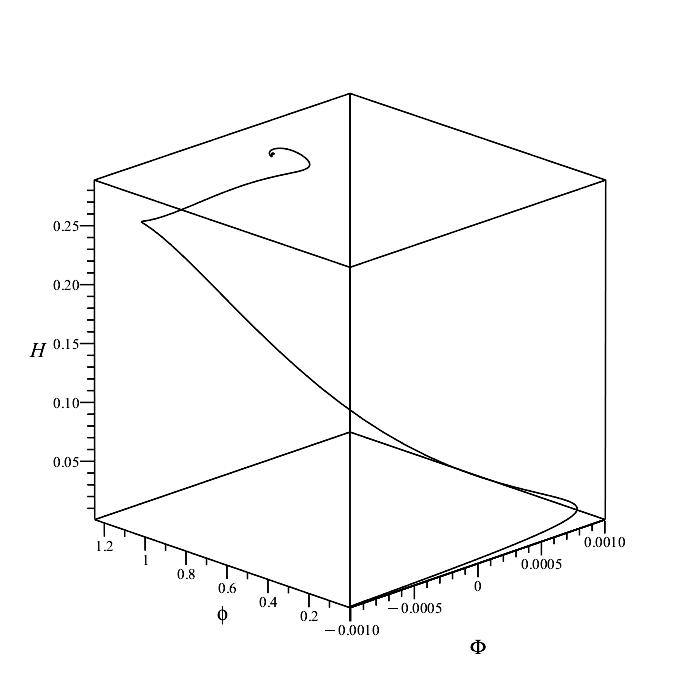}{8}{\label{fig5} Phase trajectory of the $\mathbf{P^{(2)}}$ model in the $\{\Phi,\phi,H\}$ subspace.}
\Fig{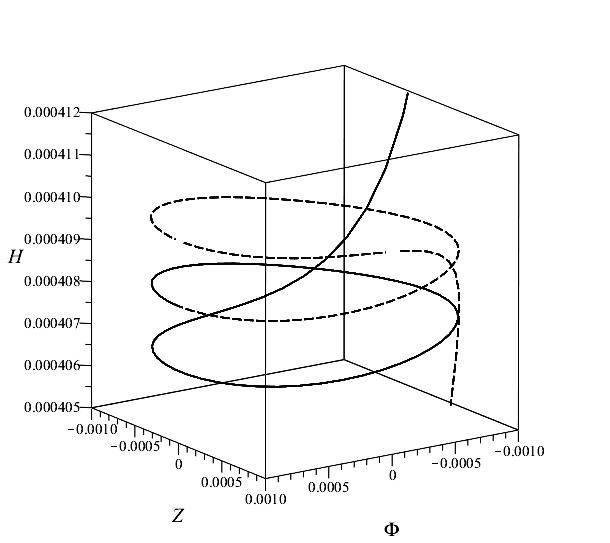}{8}{\label{fig6}Phase trajectory of the model $\mathbf{P^{(2)}}$ in the subspace $\{\Phi,Z,H\}$.}

Finally, the graph in Fig. \ref{fig6} shows the formation of an ``almost Eulidean cycle'', corresponding to the horizontal sections of the graphs in Fig. \ref{fig2}. Thus, although the assumption about the possibility of the existence of Euclidean cycles, formulated in the work \cite{Yu_Sam_Izv19}, was based on a not entirely correct mathematical model, it turned out to be correct. At this stage, almost complete compensation of the total energy of the scalar fields ASD(P) occurs.

\subsection{The influence of the interaction\newline constant $\gamma$ on the properties\newline of the cosmological model}
We consider models with a zero value of the cosmo\-lo\-gical constant $\Lambda$ and different values of the inte\-rac\-tion constant of the components $\gamma$
\begin{eqnarray} \label{par_gamma}
\mathbf{P^{(\gamma)}}&=&[[1,1,1,1],0,\gamma],
\end{eqnarray}
where $\gamma=0.49;0.6;1$
and initial conditions \eqref{ic3}
\begin{eqnarray}	\label{ic3}
\mathbf{I^{(3)}_\pm}=[0.8999550034,0,0.001,0,\pm 1].
\end{eqnarray}

Note that according to \eqref{M00} -- \eqref{Mpmpm}, the value of the constant of interaction of components $\gamma$ does not affect the coordinates of the singular points $M^\pm_{0,0}$, $M^\pm_{\pm,0}$, $M^\pm_{0,\pm}$, but only the coordinates of the singular points $M^\pm_{\pm,\pm}$, of which, according to Table \ref{tab1}, only the points $M^{+}_{\pm,\pm}$ are conditionally stable in the subspace $\Sigma_\phi$.

Table \ref{tab4} shows the nature of the singular points $M^{\pm}_{0,0}$, $M^\pm_{0,\pm}$ and $M^\pm_{\pm,0}$
\begin{center}
	\refstepcounter{table}
	{\bf Tab \thetable.} \label{TabPoint}\hskip 12pt Special points for the model \eqref{par_gamma} \\[12pt]
	\begin{tabular}[c]{|c|c|c|}
\hline
Point & Coordinates & Type\\ \hline
$M_{0,0}^{+}$ &[0, 0, 0, 0,0.182]& [\textbf{A}, \textbf{S}] \\ \hline
$M_{0,0}^{-}$ & [0, 0, 0, 0, -0.182]& [\textbf{R}, \textbf{S}] \\ \hline
$M_{\pm,0}^{+}$ &[$\pm$ 0.9, 0, 0, 0, 0.297]& [\textbf{S}, \textbf{S}] \\ \hline
$M_{\pm,0}^{-}$ &[$\pm$ 0.9, 0, 0, 0, -0.297] & [\textbf{S}, \textbf{S}] \\ \hline
$M_{0,\pm}^{+}$ &[0, 0, $\pm$ 0.9, 0, 0.297]& [\textbf{A}, \textbf{A}] \\ \hline
$M_{0,\pm}^{-}$ &[0, 0, $\pm$ 0.9, 0, -0.297] & [\textbf{R}, \textbf{R}] \\ \hline
\end{tabular}
\label{tab4}
\end{center}
Note that for the other parameters adopted in the case $\gamma>0.49$, the singular points $M^{\pm}_{\pm,\pm}$ are absent in the model.

Fig. \ref{fig7} shows the evolution of the geometric factors $\xi(t)$ and $H(t)$ for the model with parameters \eqref{par_gamma} and initial conditions \eqref{ic3}, and Fig. \ref{fig8} shows the evolution of the potential of the classical and phantom fields for this model.
Note that for $\gamma>0.49$ the models have a finite future: for $\gamma=0.6$ -- $t_\infty\approx 54.0$ and for $\gamma=1$ -- $t_\infty\approx 27.3$.

\Fig{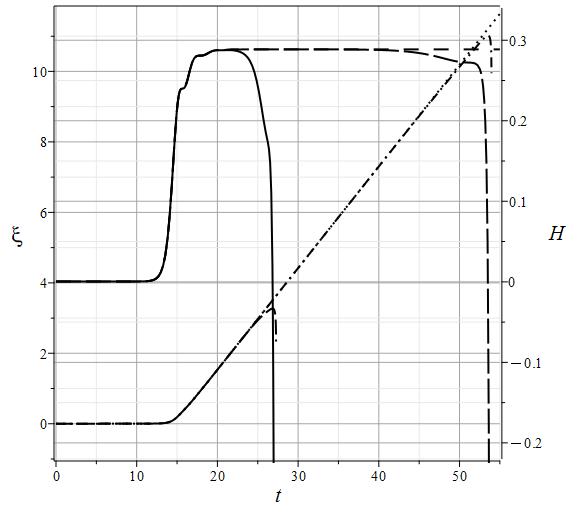}{6}{\label{fig7} $\xi(t)$: dashed line -- $\gamma=1$; dash-dotted -- $\gamma=0.6$; dotted -- $\gamma=0.49$ and $H(t)$: solid line -- $\gamma=1$; long-dashed -- $\gamma=0.6$; long-dotted -- $\gamma=0.49$; parameters $\mathbf{P^{(\gamma)}}$ and initial conditions $\mathbf{I^{(3)}}$.}
\Fig{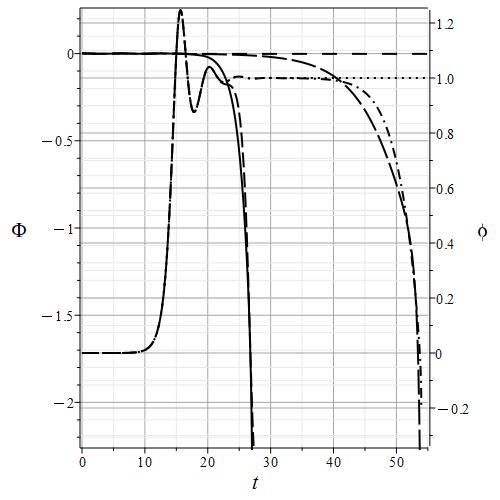}{6}{\label{fig8} $\Phi(t)$: solid line -- $\gamma=1$; long-dashed -- $\gamma=0.6$; long-dashed line -- $\gamma=0.49$ and $\phi(t)$: dashed line -- $\gamma=1$; dash-dotted -- $\gamma=0.6$; dotted -- $\gamma=0.49$; parameters $\mathbf{P^{(\gamma)}}$ and initial conditions $\mathbf{I^{(3)}}$.}

Let us consider in particular the case of a critical value of the interaction constant $\gamma=0.49$
\begin{eqnarray}	\label{par4}
\mathbf{P^{(4)}}&=&[[1,1,1,1],0.1,0.49],
\end{eqnarray}
with initial conditions \eqref{ic4}.
\begin{eqnarray} \label{ic4}
\mathbf{I^{(4)}}=[0.001,0,0.000001,0,1].
\end{eqnarray}
In this case, the coordinates of the singular points $M^{+}_{\pm,\pm}$ are $[\Phi,\phi,H]=[x,x,h]$, where $x\approx\pm 0.71$ and $h\approx 0.343$, and the signs of $x$ and $h$ take independent values.

Fig. \ref{fig9} shows the evolution of the geometric factors $\xi(t)$ and $H(t)$ for the model with parameters \eqref{par4} and initial conditions \eqref{ic4}. In this case, the model has an infinite past and an infinite future and, in addition, a rebound point $t_b\approx -11.25$.
\Fig{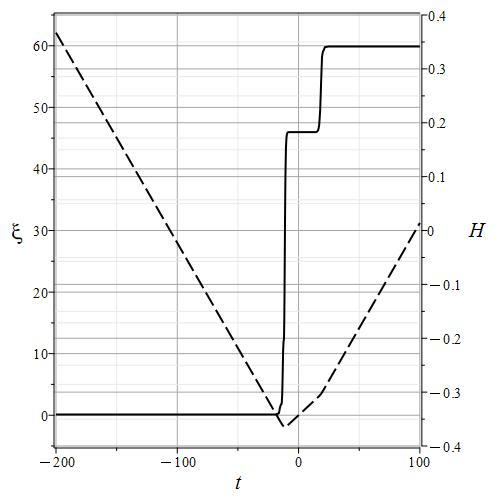}{6}{\label{fig9} $\xi(t)$ (dashed line) and $H(t)$ (solid line); parameters $\mathbf{P^{(4)}}$ \eqref{par4} and initial conditions $\mathbf{I^{(4)}}$ \eqref{ic4}.}

The small-scale picture of the evolution of scalar fields for this case is shown in Fig. \ref{fig10}, and the large-scale picture is shown in Fig. \ref{fig11}. As can be seen from these figures, near the rebound point, the process of oscillation of the potentials of these fields occurs, after which the system returns to a stable state.
\Fig{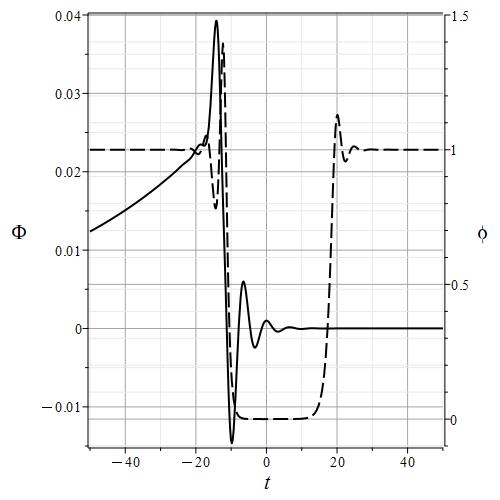}{6}{\label{fig10} $\Phi(t)$ (solid line) and $\phi(t)$ (dashed) near the rebound point $t_b$; parameters $\mathbf{P^{(4)}}$ \eqref{par4} and initial conditions $\mathbf{I^{(4)}}$ \eqref{ic4}.}
\Fig{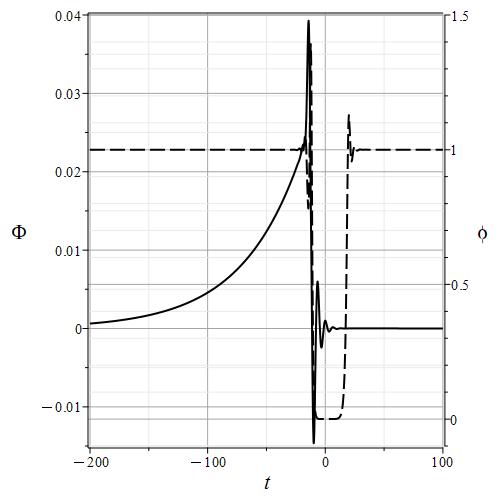}{6}{\label{fig11}$\Phi(t)$ (solid line) and $\phi(t)$ (dashed); parameters $\mathbf{P^{(4)}}$ \eqref{par4} and initial conditions $\mathbf{I^{(4)}}$ \eqref{ic4}.}

Finally, Fig. \ref{fig12} demonstrates the asymptotic properties of the system's behavior near the singu\-la\-rity for the parameters \eqref{par5} and \eqref{par6} corresponding to the model with a finite future, confirming the analytical estimates \eqref{dotH=xi}.
\begin{eqnarray} \label{par5} \mathbf{P^{(5)}}&=&[[1,1,1,1],0.1,0.6], \end{eqnarray}
\begin{eqnarray} \label{par6} \mathbf{P^{(6)}}&=&[[1,1,1,1],0.1,1],
\end{eqnarray}
\Fig{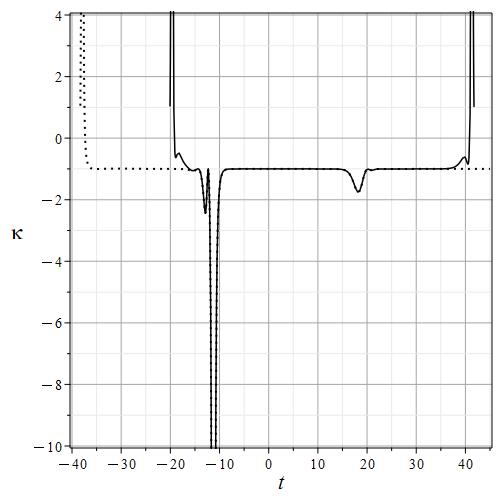}{6}{\label{fig12} Evolution of the barotropic coefficient $\kappa$: $\gamma=0.6$ (dashed line) and $\gamma=1$ (solid line) in the model with parameters $\mathbf{P^{(5)}}$ \eqref{par5}, $\mathbf{P^{(6)}}$ \eqref{par6} and initial conditions $\mathbf{I^{(4)}}$ \eqref{ic4}.}

\section{Conclusion}
Thus, the conducted study revealed some interesting features of cosmological models based on an asymmetric scalar doublet with potential interaction of the doublet components.
\begin{enumerate}
\item Depending on the values of the fundamental parameters of the ASD(P) model, a wide range of its behavior types is possible, including models with infinite inflationary past
and future, models with an initial singularity and infinite inflationary future, with an infinite inflationary past and finite future (Big Rip), with finite past and future.
\item In particular, for sufficiently large values of the interaction constant $\gamma\gtrsim0.49$, cosmological evolution quickly ends with Big Rip, which makes ASD(P) cosmological models with large values of the interaction constant unsuitable.
\item In some cases, the models have rebound points, at which the process of transition from one stable state to another occurs. This process is accompanied by strong oscillations
of scalar fields, which can, as in the ASD(K) models, lead to the birth of scalar charged fermions \cite{Yu_Kokh_24_2}.
\item In some cases, the models allow for a phase of a nearly Euclidean cycle, in which the Universe is nearly Euclidean due to the almost complete compensation of the energy of the doublet fields.
\end{enumerate}

The identified properties of the model allow us to consider it as an alternative when constructing a theory of the formation of supermassive black holes in the early Universe.

\subsection*{Funding}
The work was carried out using funds from a subsidy allocated as part of state support for Kazan (Volga Region) Federal University in order to increase its competitiveness among the world's leading scientific and educational centers.

\end{document}

%% file: preamble.tex
\usepackage{amsfonts,amssymb,cite}
\usepackage{graphicx}

\def\noi{\noindent}

\newcommand{\Title}[1]{\noi {{\Large\bf #1}}\\[1ex]}

\newcommand{\Author}[2]{\noi{\bf #1}\\[2ex]\noi{\normalsize\it #2}\\}

\newcommand{\Abstract}[1]{\vskip 2mm \begin{center}
        \parbox{16.4cm}{\small\noi #1} \end{center}\medskip}

\def\nqq{\hspace*{-2em}}

\usepackage{color}

\def\Jl#1#2{#1 {\bf #2},\ }

\def\ApJ#1 {\Jl{Astroph. J.}{#1}}
\def\CQG#1 {\Jl{Class. Quantum Grav.}{#1}}
\def\DAN#1 {\Jl{Dokl. AN SSSR}{#1}}
\def\GC#1 {\Jl{Grav. Cosmol.}{#1}}
\def\GRG#1 {\Jl{Gen. Rel. Grav.}{#1}}
\def\IJMPD#1 {\Jl{Int. J. Mod. Phys. D}{#1}}
\def\JETF#1 {\Jl{Zh. Eksp. Teor. Fiz.}{#1}}
\def\JETP#1 {\Jl{Sov. Phys. JETP}{#1}}
\def\JHEP#1 {\Jl{JHEP}{#1}}
\def\JMP#1 {\Jl{J. Math. Phys.}{#1}}
\def\NPB#1 {\Jl{Nucl. Phys. B}{#1}}
\def\NP#1 {\Jl{Nucl. Phys.}{#1}}
\def\PLA#1 {\Jl{Phys. Lett. A}{#1}}
\def\PLB#1 {\Jl{Phys. Lett. B}{#1}}
\def\PRD#1 {\Jl{Phys. Rev. D}{#1}}
\def\PRL#1 {\Jl{Phys. Rev. Lett.}{#1}}

\def\lal{&&\nqq {}}

\def\beq{\begin{equation}}
\def\eeq{\end{equation}}
\def\bear{\begin{eqnarray}}
\def\bearr{\begin{eqnarray} \lal}
\def\ear{\end{eqnarray}}
\def\earn{\nonumber \end{eqnarray}}

